\documentclass[lettersize,journal]{IEEEtran}
\usepackage{amsmath,amsfonts}
\usepackage{algorithmic}
\usepackage{algorithm}
\usepackage{array}
\usepackage{textcomp}
\usepackage{stfloats}
\usepackage{url}
\usepackage{verbatim}
\usepackage{graphicx}
\usepackage{cite}
\usepackage{xcolor}
\usepackage{booktabs}
\usepackage{pifont}
\newcommand{\cmark}{\ding{51}}  
\newcommand{\xmark}{\ding{55}}  
\usepackage{hyperref}
\usepackage{algorithm}
\usepackage{amsmath,amssymb,amsfonts}
\usepackage{algorithmic}
\usepackage{tcolorbox} 
\usepackage{multirow}
\usepackage{subcaption}
\begin{document}

\title{Contrastive Predictive Coding with Compression for Enhanced Channel State Feedback in Wireless Networks} 

\author{
	\IEEEauthorblockN{Ahmed Y. Radwan,  Fahad Syed Muhammad, Matthew Baker, and Hina~Tabassum,~\IEEEmembership{Senior~Member,~IEEE}	\thanks{A. Y. Radwan and H. Tabassum are with the Department of Electrical Engineering and Computer Science, York University, Toronto, ON M3J 1P3, Canada.  Emails: \{hinat, ahmedyra\}@yorku.ca. F. Muhammad is with Nokia Networks, 12 rue Jean Bart, Paris-Saclay, Massy, 91300, France. M. Baker is with the Nokia UK, Broers Building, 21 JJ Thomson Avenue, Cambridge, CB3 0FA, United Kingdom. Emails: \{fahad.syed\_muhammad, matthew.baker\}@nokia.com.
  This work was supported by Natural Sciences and Engineering Research Council of Canada (NSERC) CREATE grant.
    }\\
	}
}


\maketitle
\raggedbottom
\begin{abstract}
Accurate and timely channel state information (CSI) is essential for next-generation wireless systems, yet existing works treat CSI compression and CSI prediction as separate problems, both in academia and in current 3GPP studies. Consequently, channel aging remains insufficiently addressed within standardized CSI feedback pipelines.
In this article, we propose a unified compression–prediction framework that integrates Contrastive Predictive Coding (CPC) directly into the 3GPP-compliant CSI compression architecture. Instead of predicting high-dimensional CSI matrices, our approach forecasts future latent representations and jointly optimizes reconstruction fidelity and temporal predictive coherence via a combined 1-SGCS and InfoNCE objective. This design enables temporal representation learning without increasing feedback overhead.
We present two variants: CPC-before-Compression, which performs autoregressive modeling on encoded features prior to quantization, and CPC-after-Compression, which shifts temporal modeling to the base-station to reduce the  complexity of users' devices. Evaluations on 3GPP-compliant datasets from Nokia, Oppo, and CATT show that CPC-before-Compression achieves over 90\% reconstruction accuracy with 32× lower decoder GFLOPs than the 3GPP baseline, while CPC-after-Compression preserves an identical encoder footprint and the same 64-bit feedback overhead.
By unifying compression and prediction within a standardized pipeline, the proposed framework provides an age-aware, computationally efficient CSI feedback solution. 

The source code is publicly available at: https://github.com/AhmedRadwan02/cpc-3gpp
\end{abstract}

\raggedbottom

\section{Introduction}
\IEEEPARstart{C}{hannel} state information (CSI) characterizes the wireless propagation conditions between transmitters and receivers, and is a fundamental enabler of network resource management mechanisms, such as beamforming, user scheduling, and link adaptation. However, the inherent time-varying nature of wireless channels, compounded by CSI acquisition delays, leads to \textit{channel aging}, wherein CSI becomes outdated before it can be effectively utilized. This challenge is further exacerbated in 5G and beyond, where massive antenna arrays, highly directional beams, mobile transceivers, and operation at higher frequencies significantly reduce channel coherence time \cite{10643015, 10143226}. Consequently, predicting accurate  CSI  is becoming critical to sustain reliable performance in  wireless networks.
Recently, the 3rd Generation Partnership Project (3GPP)\footnote{3GPP is a global collaboration of telecom standards organizations that develops unified specifications for mobile networks (from 3G to 5G and beyond), ensuring worldwide interoperability and innovation in wireless communications.} has taken a leading role in formalizing the integration of artificial intelligence (AI) into wireless networks, with  emphasis on improving CSI feedback latency in multi-user multiple-input multiple-output (MU-MIMO) systems{~\cite{3gppTR38843_v19_0_0}}. The need for CSI prediction becomes even more evident in high-mobility environments, where the CSI varies rapidly \cite{xu2024learning}.

Accordingly, 3GPP Release-18 introduces two key AI-driven use-cases for CSI feedback enhancement: (1)\textit{ CSI compression and reconstruction}, where user equipment (UE) employs an encoder to compress CSI for efficient uplink transmission and accurate reconstruction at the base station (BS)~\cite{csi_overhead_Reduc}; and (2) \textit{CSI prediction}, where the UE forecasts future channel states to mitigate CSI aging effects {~\cite{3gppTR38843_v19_0_0}}.
The corresponding research evolution is thus reflected in 3GPP 
standardization efforts, where early AI-based CSI feedback approaches were unsatisfactory compared to 
traditional methods~\cite{sun2024combination}, but recent evaluations report up to ~50\% overhead 
reduction versus legacy methods~\cite{csi_overhead_Reduc}.

{While CSI compression has shown promising progress in the current state-of-the-art \cite{liu2020efficient, mashhadi2020distributed, guo2024deep, shao2024quantized, an2025deep, zhang2024continuous}, CSI prediction remains fundamentally limited. 
A couple of research works focus on direct time-series forecasting of raw CSI using long short-term memory 
(LSTM)~\cite{luo2018channel} and, more recently, Transformers~\cite{zhou2024transformer}, which 
rely heavily on labeled data and struggle to capture long-term dependencies. Although Transformers 
improve upon LSTMs' sequential processing constraints through attention mechanisms, both methods 
primarily predict sequence values rather than learning meaningful representations of underlying 
channel dynamics. {A survey of traditional CSI prediction methods, such as auto-regressive and parametric models, and AI-based methods covering eigenvector and channel matrix prediction. has been provided in \cite{chengyong2025ai}.}

{In summary, existing research works \emph{consider  CSI \textit{compression}} \cite{liu2020efficient, mashhadi2020distributed, guo2024deep, shao2024quantized, an2025deep, zhang2024continuous} and \textit{CSI  prediction} \cite{luo2018channel, zhou2024transformer} as  separate problems. This is also the case in current 3GPP studies, where ``CSI feedback enhancement'' has been discussed via distinct sub-use cases (e.g., CSI compression using a two-side model versus CSI prediction using a UE-side model) \cite{3gppTR38843_v19_0_0,chengyong2025ai}.}
{Recently,~\cite{kadambar2023deep} considered joint compression and prediction using a Bi-LSTM, yet operates directly on high-dimensional raw CSI vectors rather than compact latent representations, and relies solely on cosine similarity rather than jointly optimizing reconstruction fidelity and temporal predictive coherence.}

{ Different from the existing research works, we propose a framework that integrates contrastive predictive coding (CPC) into the 3GPP compression pipeline. The framework  jointly addresses CSI compression and channel aging through temporal representation learning.
That is, rather than optimizing compression fidelity alone, we jointly optimize reconstruction quality and temporal predictive coherence through a combined 1-SGCS and InfoNCE loss. Instead of predicting the CSI matrix itself, we predict future latent representations, avoiding high-dimensional CSI processing and integrating temporal prediction 
directly into the compression pipeline.
In this context, the key contributions of this article can be listed as follows:

$\bullet$ We conduct a review of AI-driven CSI feedback methods and classify them under the 3GPP training paradigms: Type-1 (single-side joint training), Type-2 (cross-side joint training), and Type-3 (separate-side training) as shown in Fig.~\ref{fig:training_types}. 
    
$\bullet$ We develop a joint compression–prediction framework that integrates CPC into a standardized CSI feedback pipeline. Unlike conventional CSI compression schemes that focus solely on reconstruction, our design incorporates CPC within a 3GPP-compliant architecture featuring a quantized linear bottleneck. The proposed approach jointly optimizes reconstruction fidelity and temporal prediction under rate constraints, addressing a challenge not considered in prior works.
    
$\bullet$ We propose  two architecturally distinct variants: \textit{CPC-before-Compression}, 
    which employs a GRU-based autoregressive module on raw encoded features before 
    dimensionality reduction, and \textit{CPC-after-Compression}, which defers temporal 
    modeling to the BS decoder, reducing UE-side complexity. Both variants learn robust 
   representations that remain stable across prediction horizons.  
    
$\bullet$ We evaluate our methods on 3GPP-compliant datasets from Nokia, Oppo, and 
    CATT, demonstrating $32\times$ lower decoder GFLOPs for \textit{CPC-before-Compression} 
    and an identical encoder footprint for \textit{CPC-after-Compression} compared to 
    the 3GPP baseline, while \textit{CPC-before-Compression} consistently achieves 
    reconstruction accuracy exceeding 90\%. Also, model compression techniques like pruning and low-rank 
factorization~\cite{cheng2017survey,han2015learning} are implemented to reduce inference costs.
}

\begin{figure*}
    \centering
    \includegraphics[width=0.69\linewidth]{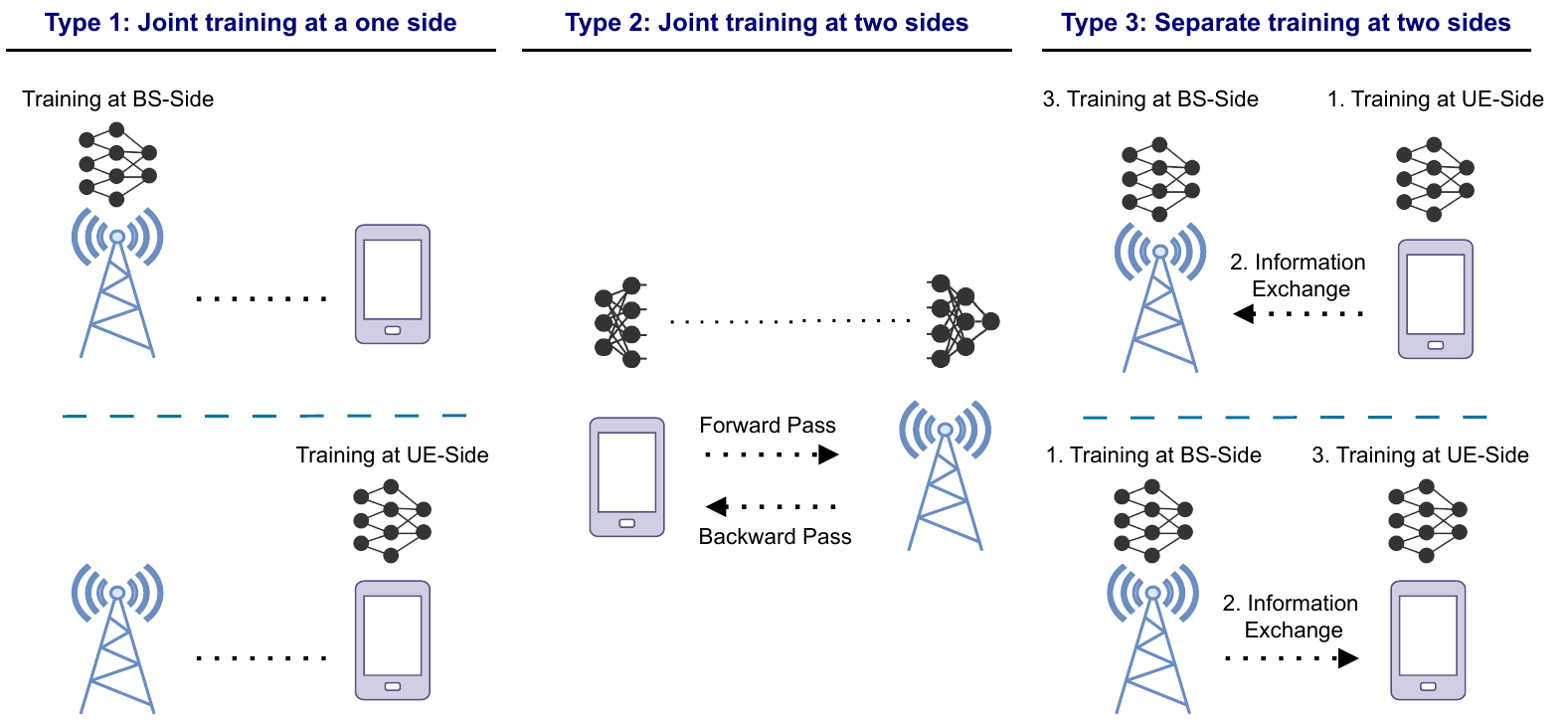}
    \caption{Overview of the model training types studied in 3GPP- Release 18: (1) Joint training of the encoder and decoder at a single side (BS or UE), (2) Joint training across two sides with intermediate activation and gradient exchange, and (3) Independent training at each side with shared datasets. The forward and backward pass flow is illustrated for each configuration.}
    
    \label{fig:training_types}
\end{figure*}

\begin{table*}[t]
\centering
\renewcommand{\arraystretch}{0.8}
\caption{Summary of Related Works on CSI Feedback Using AI/ML Techniques}
\resizebox{\textwidth}{!}{%
\begin{tabular}{lccp{2cm}ccp{4cm}p{3cm}p{2cm}}
\toprule
\textbf{Paper} & \textbf{Type of Training} & \textbf{Training Mode} & \textbf{Architecture} & \textbf{Prediction} & \textbf{Generalization} & \textbf{Datasets} & \textbf{Evaluation Metrics} & \textbf{Loss Function} \\ 
\midrule
\cite{liu2020efficient} & \textit{Type-1} (BS) & Offline & Conv Autoencoder + Quantizer
 & \xmark & \xmark & Simulated using COST 2100 model in indoor and outdoor scenarios & NMSE & MSE + Quantization regularizer  \\
\cmidrule(lr){2-9}

\cite{mashhadi2020distributed} & \textit{Type-1} (BS) & Offline & Full-Conv Autoencoder & \xmark & \xmark & Simulated using COST 2100 model in indoor and outdoor scenarios & NMSE, Cosine Correlation & MSE \\
\cmidrule(lr){2-9}

\cite{rizzello2021learning} & \textit{Type-1} (BS) & Offline & Conv Autoencoder & \xmark & \cmark & Simulated using MATLAB in an urban microcell scenario & NMSE, Cosine Similarity, Per-User Rate & MSE  \\
\cmidrule(lr){2-9}

\cite{huang2023deep} & \textit{Type-1} (BS) & Offline & Conv Variational Autoencoder & \xmark & \xmark & CSiNet Indoor Dataset & NMSE &(MSE or SURE) + MI upper-bound regularizer \\
\cmidrule(lr){2-9}

\cite{guo2024deepMiMO} & \textit{Type-1} (BS) & Offline & Conv Autoencoder & \xmark & \xmark & Simulated using COST 2100 model in indoor and outdoor scenarios & NMSE & MSE \\
\cmidrule(lr){2-9}

\cite{shao2024quantized} & \textit{Type-1} (BS) & Offline & Conv Encoder + Transformer Decoder & \xmark & \xmark & Simulated using Quadriga v2.8.1 in 3GPP-38.901 UMi and UMa & NMSE & MSE + Quantization Loss (codebook + encoder terms)\\
\cmidrule(lr){2-9}

\cite{an2025deep} & \textit{Type-1} (BS) & Offline & & \xmark & \xmark & Simulated using COST 2100 model in indoor and outdoor scenarios & NMSE & \\
\cmidrule(lr){2-9}

\cite{zhang2024continuous} & \textit{Type-1} (UE) & Online & Conv Autoencoder & \xmark & \xmark & Simulated using CDL model in 3GPP TR 38.901 & NMSE & MSE + Regularizer \\
\cmidrule(lr){2-9}

\cite{saini2024network} & \textit{Type-3} & Offline & Conv Autoencoder & \xmark & \xmark & Simulated using 3GPP TR 38.901 UMa scenario & NMSE, SGCS, SE, Overhead & SGCS + NMSE  \\
\cmidrule(lr){2-9}

\cite{tan2024federated} & \textit{Type-3} & Online & Conv Autoencoder & \xmark & \cmark & Simulated using COST 2100 and CDL from 3GPP TR 38.901 R15 & NMSE, Overhead & MSE + Regularizer \\
\cmidrule(lr){2-9}

\textbf{Ours} & \textbf{\textit{Type-1 (BS)}} & \textbf{Offline} & \textbf{Conv Autoencoder + CPC} & \textbf{\cmark} & \textbf{\cmark} & \textbf{3GPP-compliant datasets (Nokia, Oppo, CATT)~\cite{3gpp_ran4_r4_113}} & \textbf{SGCS, Complexity, Inference Time, Overhead} & \textbf{InfoNCE + (1-SGCS)} \\
\bottomrule
\end{tabular}}
\footnotesize
NMSE: Normalized Mean Square Error, CDL: Clustered Delay Line, SGCS: Squared Generalized Cosine Similarity, SE: Spectral Efficiency, UMi: Urban Micro, UMa: Urban Macro, FDD: Frequency Division Duplex, MSE: Mean Square Error.
\label{tab:literature_review_summary}
\end{table*}

\section{AI-Driven CSI Feedback: A Review}

This section reviews existing AI-driven CSI feedback methods that support 3GPP  objectives.  3GPP categorizes AI-driven CSI compression and reconstruction into three distinct types~\cite{R1-2308873}, as summarized in \textbf{Table-\ref{tab:literature_review_summary}}.

$\bullet$ \textit{Type-1} involves joint training of the full model at a single side, either the BS or UE, {followed with a split deployment at UE and BS}. 
This approach is simple and resource-efficient, but lacks post-deployment adaptability, making it less effective in dynamic environments. 

{$\bullet$ \textit{Type-2} enables collaborative training between the BS and UE where the same model instance is split between UE and BS. This method enables end-to-end optimization} through the exchange of intermediate features, gradients, or activations. 
While this improves adaptability to real-time channel variations, it introduces significant communication overhead, latency, and system complexity. 

$\bullet$ \textit{Type-3} refers to separate training of {different models}  at each side. 
This setup, which may also include federated or distributed learning, offers modularity and improved generalization, but poses challenges in synchronization, cross-device compatibility, and computational efficiency at the UE. 

A graphical illustration of the model training types studied in 3GPP Rel-18 is given in Fig.~\ref{fig:training_types}.

Current research works are predominantly based on Type-1 training due to its implementation simplicity and deployment practicality. However,  a growing number of studies have begun exploring Type-3 training recently, motivated by its potential for improved generalization in heterogeneous wireless environments. Notable works leveraging Type-1 training include~\cite{rizzello2021learning,huang2023deep,guo2024deepMiMO,shao2024quantized,an2025deep,zhang2024continuous}, while emerging studies investigating Type-3 include~\cite{sun2024combination,tan2024federated, saini2024network}. The following subsections review representative works in both categories.

\subsection{Type-1: Joint Training at One Side}

{Early CSI compression efforts, such as~\cite{rizzello2021learning}, demonstrated that an autoencoder trained on uplink CSI at the BS could generalize across frequencies. The encoder can then be offloaded to the UE for CSI feedback, while the decoder remains at the BS.}
Building on those principles,~\cite{huang2023deep} introduced a two-part training objective to improve reconstruction without requiring access to clean training data. The first leverages Stein’s Unbiased Risk Estimator (SURE) to optimize reconstruction from noisy CSI, while the second adds a compression regularization term (derived via a variational upper bound on mutual information) to encourage a compact latent representation. Together, these terms balance compression efficiency and reconstruction fidelity. 
The authors in \cite{guo2024deepMiMO} applied deep learning to jointly address channel estimation and feedback in Frequency Division Duplex (FDD) massive MIMO systems. Their primary model, CEFnet, consists of four main stages: CSI channel estimation using pilots, a feature extraction network to process the estimated channel, an encoder to perform dimensionality reduction, and a quantization module to enable efficient transmission. On the BS side, an additional refinement stage is introduced to enhance the decompressed CSI using CNN. {To address the computational limitations of the UE, the authors also proposed an alternative model, CEFnet-B, which skips channel estimation at the UE. Instead, raw pilot signals are directly compressed and quantized, and after decompression at the BS, CSI is estimated and subsequently refined.} This design shifts the computational burden entirely to the BS.


Similarly, in \cite{shao2024quantized}, {the authors considered more computations at the BS side.} The architecture employs a lightweight encoder for CSI compression followed by the quantization layer at UE. In contrast, the decoder at the BS is more complex, as after dequantization a transformer-based network equipped with multi-head attention and positional encoding has been applied to learn interactions across distant antenna and subcarrier elements to effectively reconstruct the CSI. End-to-end offline training was performed  at the BS, after which the encoder and quantization layers were transferred to the UE. 

{\cite{liu2020efficient} proposed CQNet, which inserts a learnable quantizer between encoder and decoder to jointly optimize compression and quantization. While this work represents an important step in learned CSI compression, it  operates entirely on static CSI snapshots and does not model temporal dependencies or mitigate channel aging.}
{In~\cite{mashhadi2020distributed}, the authors proposed DeepCMC, a fully convolutional autoencoder with entropy coding that approaches the rate-distortion bound for CSI feedback. Similarly, this work focuses exclusively on compression fidelity, with no mechanism for temporal prediction or robustness to CSI aging.
}

In~\cite{an2025deep}, the authors proposed LWNet, which further reduced UE computational demands by processing CSI samples in parallel using convolutional filters with different kernel sizes to extract features at multiple granularities before merging. An Efficient Channel Attention (ECA) module enabled global pooling without dimensionality reduction, while the decoder used Omni-Dimensional Dynamic Convolutions (ODConv) to optimize computation. This design emphasized a lightweight UE encoder while offloading complexity to the BS. Lastly,~\cite{zhang2024continuous} tackled key limitations of the offline training paradigm, which is predominantly used in Type-1 settings, by incorporating an online learning mechanism directly at the UE. Their approach enabled local fine-tuning of both the encoder and decoder modules before transmitting the updated model back to the BS. {However, while using online learning improves adaptability, deploying complex decoder architectures—such as those in~\cite{shao2024quantized} may not be feasible, as it is limited by the UE’s limited computational resources. Moreover, the method did not explicitly account for the quality of incoming data; thus fine-tuning on low-quality or noisy CSI samples could inadvertently degrade reconstruction performance.}

\subsection{Type-3: Separate Training at Two Sides}
As mentioned earlier, there are only a handful of recent research works that fall under this category. Type-3 decouples encoder and decoder training between the UE and BS, primarily to enhance privacy and reduce communication overhead. \cite{saini2024network} proposed a strategy where the UE trains the encoder offline and transmits latent representations to the BS, and the decoder is trained separately—minimizing data exchange and privacy risks. {Another way to implement Type-3 is via Federated Learning (FL), which enables multiple UEs to contribute to a global model deployed on a BS by sending their local model weights or gradients.} As shown in~\cite{sun2024combination, tan2024federated}, distributed training across UEs helps preserve privacy and reduces reliance on the BS, as it aggregates local updates rather than performing centralized training. These approaches also incorporate compressed sensing, knowledge distillation, and local fine-tuning to personalize models with minimal overhead.

\begin{figure*}[t]
    \centering
    \includegraphics[width=0.79\linewidth]{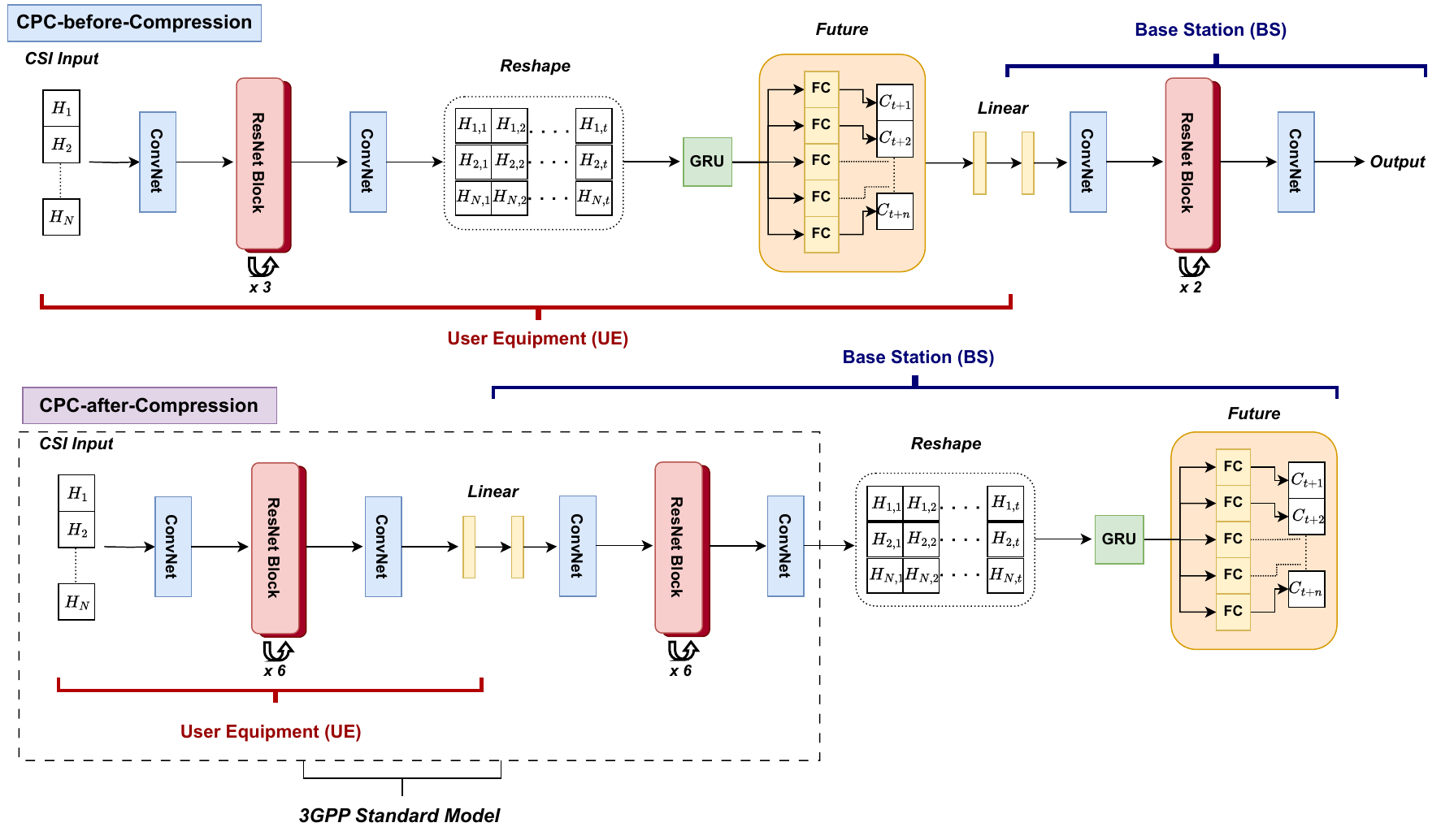}
    \caption{The CPC-after-Compression model (below) builds on the 3GPP standard backbone by adding CPC to compressed representations, while CPC-before-Compression (top) applies CPC directly on raw CSI features.}
    \label{fig:CPC_Compression}
\end{figure*}

\subsection{Key Takeaways and Summary}

Despite recent advances in AI-driven CSI feedback methods,  existing CSI compression studies~\cite{rizzello2021learning,huang2023deep,guo2024deepMiMO,shao2024quantized,an2025deep,zhang2024continuous,sun2024combination,tan2024federated, saini2024network} do not address the problem of CSI prediction jointly. Moreover, the existing CSI prediction works have neither investigated the benefits of representation learning, nor explored the use of open-access 3GPP-compliant datasets for validation. {However, a recent study in~\cite{kadambar2023deep} considered prediction directly on raw CSI vectors using a Bi-LSTM, either at the UE before compression or at the BS after decompression, with quantization applied only on the latent space. While  jointly trains compression and prediction using only cosine similarity, our CPC-before-Compression uses a combined 1-SGCS and InfoNCE loss to explicitly encourage temporal predictive structure in the latent space.} 

\section{Proposed Age-Aware CPC-based CSI Feedback}
    In this section, we propose a 3GPP-compliant AI solution that explicitly targets channel aging and supports robust generalization across deployment scenarios. Specifically, we incorporate CSI prediction through CPC~\cite{{oord2018representation}} into the 3GPP standard compression model. This integrated framework, referred to as \textit{CPC with Compression}, leverages temporal dependencies in the CSI sequence to improve both prediction robustness and compression effectiveness. By forecasting future CSI representations, the model compensates for delays between CSI acquisition and usage, which is critical in highly dynamic environments.
    On the other hand, generalization is achieved through a comprehensive evaluation across diverse 3GPP-compliant datasets, highlighting the model’s adaptability beyond the training environment. Furthermore, we incorporate structured pruning and aggressive quantization to reduce model size, enabling efficient deployment on resource-constrained UEs without compromising performance. For our training paradigm, we build upon 3GPP Type-1 training framework due to its simplicity, low deployment complexity, and compatibility with telecommunication systems. 

    
    \subsubsection{AI/ML Models Considered in 3GPP}
    \label{subsec:3GPPModel}
    
    The compression model considered in 3GPP from \cite{3gpp2024requirementsModel} employs a 
    ResNet-based architecture for end-to-end training of both encoder and decoder components. 
    The encoder begins with a convolutional layer that maps input CSI {$\mathbf{H}_t$} to 64 
    feature maps, followed by six ResNet blocks---each consisting of two convolutional layers 
    with skip connections. By concatenating the input with the block output, the architecture 
    enhances gradient flow and mitigates vanishing gradient issues. The output from the final 
    ResNet block is projected into a {32-dimensional latent vector $\mathbf{z}_t$ via a quantized 
    linear layer}, yielding a compact yet informative representation that reduces CSI feedback 
    overhead. {On the decoder side, $\mathbf{z}_t$ is dequantized and progressively reconstructed 
    back to the original CSI $\tilde{\mathbf{H}}_t$, mirroring the encoder in structure. To 
    quantify reconstruction quality, we adopt the Squared Generalized Cosine Similarity (SGCS) 
    metric, defined as:
    \begin{equation}
    \label{eq:SGCS}
    \text{SGCS} = \frac{1}{B} \sum_{i=1}^{B} \frac{1}{1 + 
    \frac{\sum_{d} |\mathbf{H}_i - \tilde{\mathbf{H}}_i|^2}{\sum_{d} |\mathbf{H}_i|^2 
    + \epsilon}},
    \end{equation}
    where $\mathbf{H}_i$ and $\tilde{\mathbf{H}}_i$ denote the ground-truth and reconstructed 
    CSI respectively, $B$ is the batch size, and $\epsilon = 1 \times 10^{-10}$ is added to 
    avoid division by zero. Higher SGCS values indicate stronger geometric alignment between 
    the original and reconstructed CSI. The model is then trained by minimizing:
    \begin{equation}
    \label{eq:SGCS_loss}
    \mathcal{L}_{\text{SGCS}} = 1 - \text{SGCS},
    \end{equation}
    such that minimizing $\mathcal{L}_{\text{SGCS}}$ is equivalent to maximizing the SGCS metric.}
    
    The overall design intentionally avoids complex modules such as attention mechanisms or 
    advanced preprocessing, prioritizing computational efficiency. This leads to reduced 
    inference latency and makes the model well-suited for deployment in resource-constrained 
    environments.
    
    Leveraging on models considered in 3GPP, we develop two design variants that address 
    channel aging. The variants are described in the following.

\begin{algorithm}[t]
\caption{CPC-before-Compression}
\label{alg:cpc_before}
\begin{algorithmic}[1]
\REQUIRE CSI sequence $\mathbf{H}_{t-L+1:t}$ at UE
\ENSURE Compressed feedback $\mathbf{z}_t$; reconstructed predicted representations 
$\{\hat{\mathbf{C}}_{t+k}\}_{k=1}^T$ at BS
\STATE \textbf{UE Processing:}
\STATE $\mathbf{F}_{t-L+1:t} \leftarrow \text{Encoder}_{\theta}(\mathbf{H}_{t-L+1:t})$
\STATE $\mathbf{F}_{\text{flat}} \leftarrow \text{ReshapeToSequence}(\mathbf{F}_{t-L+1:t})$
\STATE $\mathbf{C}_t \leftarrow \text{GRU}(\mathbf{F}_{\text{flat}})$ \hfill{\small (temporal modeling before compression)}
\FOR{$k = 1$ \textbf{to} $T$}
    \STATE $\mathbf{C}_{t+k} \leftarrow \text{FC}_k(\mathbf{C}_t)$
\ENDFOR
\STATE $\mathbf{z}_{t+k} \leftarrow \text{Quantize}(\text{LinearCompress}(\mathbf{C}_{t+k}))$ \hfill{\small (for each $k=1,\ldots,T$)}
\STATE \textbf{Transmit} $\{\mathbf{z}_{t+k}\}_{k=1}^T$ via uplink
\STATE \textbf{BS Processing:}
\STATE $\mathbf{C}'_{t+k} \leftarrow \text{Dequantize}(\mathbf{z}_{t+k})$ \hfill{\small (for each $k=1,\ldots,T$)}
\STATE $\hat{\mathbf{C}}_{t+k} \leftarrow \text{LinearExpand}(\mathbf{C}'_{t+k})$ \hfill{\small (for each $k=1,\ldots,T$)}
\STATE $\mathcal{L}_{\text{SGCS}} \leftarrow$ Using \eqref{eq:SGCS_loss} 
$(\{\mathbf{C}_{t+k}\}, \{\hat{\mathbf{C}}_{t+k}\})$
\STATE $\mathcal{L}_{\text{InfoNCE}} \leftarrow$ Using \eqref{eq:infonce} 
$(\mathbf{C}_t, \{\mathbf{C}_{t+k}\})$
\STATE $\mathcal{L}_{\text{total}} \leftarrow$ Using \eqref{eq:combined_loss}
\STATE \textbf{return} $\mathbf{z}_t, \{\hat{\mathbf{C}}_{t+k}\}_{k=1}^T$
\end{algorithmic}
\end{algorithm}

   \subsubsection{CPC-before-Compression}
    Applying CPC before compression allows temporal features to be encoded prior to 
    dimensionality reduction. This design facilitates easier reconstruction (as shown in 
    Fig.~\ref{fig:CPC_Compression} and Algorithm~\ref{alg:cpc_before}) because the 
    representation is already mapped using an \textit{autoregressive GRU-based model} 
    \cite{chung2014empirical} as a predictive layer to capture temporal dependencies and 
    predict future CSI representations. The architecture comprises an encoder, a predictive 
    module, a compression layer, and an optional decoder since the representation is restored 
    from the first layer. However, since most computations occur on the encoder side, this 
    design requires more resources, leaving less capacity for optimization on resource-limited 
    UE devices.
    
    {The encoder takes as input a sequence of $L$ consecutive CSI matrices 
    $\mathbf{H}_{t-L+1:t}$, each mapped to 64 feature maps via an initial convolutional layer, 
    followed by three ResNet blocks and a final $1{\times}1$ convolution, producing the feature 
    sequence $\mathbf{F}_{t-L+1:t}$. The features are then reshaped into a temporal sequence 
    $\mathbf{F}_{\text{flat}}$ and passed to a GRU module, which summarizes the temporal 
    context into a 128-dimensional context vector $\mathbf{C}_t$. A set of $T$ dedicated 
    fully-connected layers then projects $\mathbf{C}_t$ into future predicted representations 
    $\{\mathbf{C}_{t+k}\}_{k=1}^{T}$, each of 128 dimensions. The predicted representations 
    are subsequently compressed into a 32-dimensional latent vector $\mathbf{z}_t$ via a 
    quantized linear layer and transmitted to the BS. At the BS, $\mathbf{z}_t$ is dequantized 
    and expanded back to 128 dimensions via a linear layer, followed by two residual blocks, 
    yielding the reconstructed predicted representations $\{\hat{\mathbf{C}}_{t+k}\}_{k=1}^{T}$.}
    
    {To train the model, we introduce the InfoNCE loss as given below
    \cite{oord2018representation}. InfoNCE loss encourages $\mathbf{C}_t$ to be predictive of future 
    representations in the latent space:
    \begin{align}
\label{eq:infonce}
\mathcal{L}_{\text{InfoNCE}} 
&= \frac{1}{T} \sum_{k=1}^{T} \ell_k, \\
\ell_k 
&= - \frac{1}{B} \sum_{i=1}^{B} 
\log 
\frac{\exp\!\left( \frac{\mathbf{C}_{t+k}^{(i)} \cdot \mathbf{C}_t^{(i)}}{\tau} \right)}
{\sum_{j=1}^{B} 
\exp\!\left( \frac{\mathbf{C}_{t+k}^{(i)} \cdot \mathbf{C}_t^{(j)}}{\tau} \right)} .
\end{align}
    where $T$ is the number of future time steps, $\tau$ is the temperature parameter, and 
    $B$ is the batch size. The SGCS loss is computed between the original predicted 
    representations $\{\mathbf{C}_{t+k}\}$ and their reconstructions 
    $\{\hat{\mathbf{C}}_{t+k}\}$. The total training loss is then given by:
    \begin{equation}
    \label{eq:combined_loss}
    \mathcal{L}_{\text{total}} = \alpha \mathcal{L}_{\text{SGCS}} + (1 - \alpha) 
    \mathcal{L}_{\text{InfoNCE}},
    \end{equation}
    where $\alpha \in [0, 1]$ controls the trade-off between reconstruction quality and temporal prediction accuracy. In our experiments, $\alpha = 0.5$ was selected empirically, as equal weighting was found to balance reconstruction fidelity and contrastive learning stability across all datasets.}

    \begin{algorithm}[t]
    \caption{CPC-after-Compression}
    \label{alg:cpc_after}
    \begin{algorithmic}[1]
    \REQUIRE CSI matrix $\mathbf{H}_t$ at UE
    \ENSURE Compressed feedback $\mathbf{z}_t$; reconstructed CSI $\widetilde{\mathbf{H}}_t$ 
    at BS; future representations $\{\mathbf{C}_{t+k}\}_{k=1}^T$
    \STATE \textbf{UE Processing:}
    \STATE $\mathbf{F}_t \leftarrow \text{Encoder}_{\theta}(\mathbf{H}_t)$
    \STATE $\mathbf{z}_t \leftarrow \text{Quantize}(\text{LinearCompress}(\mathbf{F}_t))$
    \STATE \textbf{Transmit} $\mathbf{z}_t$ via uplink
    \STATE \textbf{BS Processing:}
    \STATE $\mathbf{F}'_t \leftarrow \text{Dequantize}(\mathbf{z}_t)$
    \STATE $\widetilde{\mathbf{H}}_t \leftarrow \text{Decoder}_{\phi}(\mathbf{F}'_t)$
    \STATE $\widetilde{\mathbf{H}}_{\text{flat}} \leftarrow 
    \text{ReshapeToSequence}(\widetilde{\mathbf{H}}_{t-L+1:t})$
    \STATE $\mathbf{C}_t \leftarrow \text{GRU}(\widetilde{\mathbf{H}}_{\text{flat}})$ 
    \hfill{\small (CPC after reconstruction)}
    \FOR{$k = 1$ \textbf{to} $T$}
        \STATE $\mathbf{C}_{t+k} \leftarrow \text{FC}_k(\mathbf{C}_t)$
    \ENDFOR
    \STATE $\mathcal{L}_{\text{SGCS}} \leftarrow$ Using \eqref{eq:SGCS_loss} 
    $(\mathbf{H}_t, \widetilde{\mathbf{H}}_t)$
    \STATE $\mathcal{L}_{\text{InfoNCE}} \leftarrow$ Using \eqref{eq:infonce} 
    $(\mathbf{C}_t, \{\mathbf{C}_{t+k}\})$
    \STATE $\mathcal{L}_{\text{total}} \leftarrow$ Using \eqref{eq:combined_loss}
    \STATE \textbf{return} $\mathbf{z}_t, \widetilde{\mathbf{H}}_t, \{\mathbf{C}_{t+k}\}_{k=1}^T$
    \end{algorithmic}
    \end{algorithm}

    \subsubsection{CPC-after-Compression}
    
    Applying CPC after compression, as shown in Fig.~\ref{fig:CPC_Compression} and 
    Algorithm~\ref{alg:cpc_after}, is more suitable for resource-limited UE devices. Since 
    the predictive layer contributes most to the model's complexity, performing compression 
    first reduces the computational burden on the UE side. In this method, CPC is integrated 
    entirely into the decoder at the BS, after the final convolutional layer responsible for 
    reconstruction.
    {The encoder is identical to the 3GPP baseline, mapping each input CSI 
    matrix $\mathbf{H}_t$ to a 32-dimensional latent vector $\mathbf{z}_t$ via a convolutional 
    feature extractor followed by six ResNet blocks and a quantized linear layer. At the BS, 
    $\mathbf{z}_t$ is dequantized and decoded through six ResNet blocks back to the 
    reconstructed CSI $\tilde{\mathbf{H}}_t$. The reconstructed frames 
    $\tilde{\mathbf{H}}_{t-L+1:t}$ are then reshaped into a temporal sequence and passed to a 
    GRU module, which generates a 128-dimensional context vector $\mathbf{C}_t$. A set of $T$ 
    fully-connected layers then projects $\mathbf{C}_t$ into future predicted representations 
    $\{\mathbf{C}_{t+k}\}_{k=1}^{T}$. The model is trained using $\mathcal{L}_{\text{total}}$ 
    from \eqref{eq:combined_loss}, where $\mathcal{L}_{\text{SGCS}}$ from \eqref{eq:SGCS_loss} 
    is computed between $\mathbf{H}_t$ and $\tilde{\mathbf{H}}_t$, and 
    $\mathcal{L}_{\text{InfoNCE}}$ from \eqref{eq:infonce} is computed between $\mathbf{C}_t$ 
    and $\{\mathbf{C}_{t+k}\}_{k=1}^{T}$.}

    \subsubsection{Insights} By shifting the computational load to the BS, \textit{CPC-after-Compression} prioritizes efficiency over early feature preservation. However, this trade-off may  impact prediction accuracy compared to the \textit{CPC-before-Compression} method, as some information is lost before temporal dependencies are fully captured. Additionally, applying \textit{CPC-after-Compression} aligns better with {split learning (SL) frameworks.} {\textit{Both CPC-before and CPC-after-Compression}} help reduce computational overhead on the UE by offloading more processing tasks to the BS. However, \textit{CPC-after-Compression} becomes  more viable when the UE has strict resource constraints.

\begin{table}[t]
\centering
\scriptsize
\caption{Model parameters, storage, inference time, and computational cost (batch size = 256, 2-bit quantization).}
\label{tab:model_parameters_and_times}
\renewcommand{\arraystretch}{1.1}
\setlength{\tabcolsep}{3pt}
\begin{tabular}{lccc}
\toprule
\textbf{Metric} & \textbf{3GPP} & \textbf{CPC-before} & \textbf{CPC-after} \\
\midrule
Encoder Params 
& \begin{tabular}[c]{@{}c@{}}872,448\\(213.00 KB)\end{tabular}
& \begin{tabular}[c]{@{}c@{}}10,587,168\\(2,574.13 KB)\end{tabular}
& \begin{tabular}[c]{@{}c@{}}872,448\\(213.00 KB)\end{tabular} \\

Decoder Params 
& \begin{tabular}[c]{@{}c@{}}1,358,786\\(331.74 KB)\end{tabular}
& \begin{tabular}[c]{@{}c@{}}103,296\\(25.22 KB)\end{tabular}
& \begin{tabular}[c]{@{}c@{}}1,810,754\\(430.00 KB)\end{tabular} \\

Encoder Time (ms) 
& 2.4533 & 2.0984 & 2.4457 \\

Decoder Time (ms) 
& 2.5065 & 0.6799 & 2.9415 \\

Encoder GFLOPs 
& 0.59 & 10.47 & 0.59 \\

Decoder GFLOPs 
& 0.84 & 0.026 & 1.21 \\
\bottomrule
\end{tabular}
\end{table}

\section{Experimental Setup \& Results}
    {This section discusses the experimental set-up, datasets, and validates the performance of the proposed model over a variety of datasets. We follow the agreed-upon model hyperparameters from 3GPP drafts, summarized in Table~\ref{tab:hyperparameters}.}

 { The evaluation is performed on proprietary datasets provided by Nokia, Oppo, and CATT, specifically \texttt{N0KIR4\_113dsei.npy}, \texttt{OPPOR4\_113dsei00.npy}, and \texttt{CAT0R4\_113dsei01.npy}~\cite{3gpp_ran4_r4_113}. To ensure a broader evaluation, we construct a mixed dataset by combining 100k samples from each of these datasets. The datasets follow a four-dimensional structure spanning samples, real/imaginary components, sub-bands (13), and antenna ports (32), and vary in the number of samples contributed by different companies.

Table~\ref{tab:model_parameters_and_times} summarizes the complexity of each model variant.

$\bullet$ The \textit{3GPP baseline} maintains a compact encoder (872K parameters, 2.45~ms per batch) with moderate decoder cost. 

$\bullet$ \textit{CPC-before-Compression} incurs a significantly larger encoder footprint (10.6M parameters, ${\sim}17.8\times$ more GFLOPs per batch: 10.47 vs. 0.59~GFLOPs) due to the GRU operating on intermediate feature maps of dimension $64 \times 13 \times 32 = 26{,}624$. Despite this, encoder inference remains comparable (2.10~ms per batch) as both models fall into a memory-bandwidth-bound regime where wall-clock time is insensitive to arithmetic complexity~\cite{cuda_graphs_overhead}. Its decoder is substantially lighter (103K parameters, 0.026~GFLOPs per batch) as temporal modeling is offloaded to the encoder side.

$\bullet$ \textit{CPC-after-Compression} preserves the same encoder as the baseline (0.59~GFLOPs per batch, 2.45~ms per batch) while adding moderate decoder overhead (1.81M parameters, 1.21~GFLOPs per batch) to accommodate the BS-side GRU. 

All methods maintain a 64-bit communication overhead per timestep, as the quantized linear layer outputs 32 latent dimensions with 2-bit quantization.}  

    {In our setup, we follow the Type-1 joint training scheme described in~\cite{3gpp2024csiCompression}. For evaluation, we use the SGCS metric instead of the commonly used NMSE, as SGCS offers a more robust measure of subspace preservation and geometric alignment. Table~\ref{tab:sgcs_infonce_results} presents the SGCS  and InfoNCE loss values for each  variant.}
    
\begin{table}[t]
\centering
\caption{Hyperparameters and Their Values}
\label{tab:hyperparameters}
\setlength{\tabcolsep}{4pt}
\renewcommand{\arraystretch}{0.9}
\begin{tabular}{ll|ll}
\toprule
\textbf{Hyperparameter} & \textbf{Value} &
\textbf{Hyperparameter} & \textbf{Value} \\
\midrule
Optimizer & Adam &
Learning Rate & $10^{-4}$ \\

Number of Epochs & 150 &
Batch Size & 256 \\

Validation Patience & 25 &
Data Split & 80:10:10 \\

Number of  Seeds & 5 &
Time Window ($L$) & 10 \\

Hidden Size & 128 &
Future Time Steps ($T$) & 5 \\

Temperature ($\tau$) & 0.1 &
Contribution parameter ($\alpha$) & 0.5 \\

Uniform Quantization & 2 bits & & \\
\bottomrule
\end{tabular}
\end{table}

    \subsection{Overall Performance Evaluation Results}
    \paragraph{Performance From Considered Models in 3GPP} We begin with the models considered in 3GPP, which serves as a baseline using SGCS only. The Nokia-trained model achieves an SGCS of {0.729} on its own test set and {0.735} on Oppo, while the Oppo-trained model attains {0.733} on its own data and {0.726} on Nokia. These results suggest that Nokia's and Oppo's models share consistent characteristics. On the other hand, the CATT-trained model, despite achieving {0.674} on its own test set, shows noticeably lower generalization to other datasets (SGCS between {0.637} and {0.644}), likely due to both its smaller size (100k samples) and possible domain-specific variability. We note that training on a mixed dataset improved performance across the board, particularly for Nokia ({0.725}), Oppo ({0.732}), and Mixed ({0.691}) test sets, although generalization to CATT remained limited ({0.614}).     
    {
    \paragraph{CPC-before-Compression Performance}
    Table~\ref{tab:sgcs_infonce_results} shows that CPC-before-Compression consistently outperforms the 3GPP baseline across all training and test configurations. SGCS scores range from 0.720 to 0.925, with CATT and OPPO-trained models achieving the strongest reconstruction fidelity. The NOKIA-trained model exhibits lower SGCS and higher InfoNCE loss (e.g., 0.726 / 0.646 on Nokia test), consistent with the training instability discussed in the ablation studies. Training on the mixed dataset yields balanced generalization across all test domains (0.885--0.887), confirming robustness to domain shift. InfoNCE loss varies widely across dataset pairs (0.150--0.666), reflecting differences in temporal predictability across channel environments. To reduce model complexity, we apply post-training structured pruning to the GRU layer in CPC-before-Compression. As shown in Fig.~\ref{fig:pruning-analysis}, increasing the pruning ratio leads to a steady decline in SGCS and a sharp rise in InfoNCE loss, with performance remaining acceptable up to moderate pruning ratios before degrading significantly at higher sparsity levels.
    }
    \begin{table}[t]
\centering
\scriptsize
\caption{SGCS ($\uparrow$) and InfoNCE ($\downarrow$) across datasets.}
\label{tab:sgcs_infonce_results}
\renewcommand{\arraystretch}{1.1}
\begin{tabular}{@{}l@{\hspace{10pt}}c@{\hspace{10pt}}c@{\hspace{10pt}}c@{\hspace{10pt}}c@{}}
\toprule
\textbf{Training} & \textbf{Nokia} & \textbf{Oppo} & \textbf{CATT} & \textbf{Mixed} \\
\midrule
\multicolumn{5}{c}{\textbf{3GPP Standard (no-CPC) (SGCS only)}}\\
\midrule
NOKIA & 0.729 & 0.735 & 0.563 & 0.676 \\
OPPO  & 0.726 & 0.733 & 0.560 & 0.673 \\
CATT  & 0.637 & 0.644 & 0.674 & 0.656 \\
Mixed & 0.725 & 0.732 & 0.614 & 0.691 \\
\midrule
\multicolumn{5}{c}{\textbf{CPC-before-Compression}}\\
\midrule
NOKIA & 0.726 (0.646) & 0.725 (0.639) & 0.738 (0.668) & 0.720 (0.664) \\
OPPO  & 0.869 (0.150) & 0.869 (0.147) & 0.866 (0.338) & 0.867 (0.158) \\
CATT  & 0.912 (0.666) & 0.912 (0.651) & 0.925 (0.279) & 0.917 (0.313) \\
Mixed & 0.886 (0.186) & 0.886 (0.182) & 0.887 (0.260) & 0.885 (0.183) \\
\midrule
\multicolumn{5}{c}{\textbf{CPC-after-Compression}}\\
\midrule
NOKIA & 0.730 (0.008) & 0.735 (0.008) & 0.566 (0.074) & 0.677 (0.009) \\
OPPO  & 0.733 (0.008) & 0.739 (0.008) & 0.568 (0.068) & 0.680 (0.009) \\
CATT  & 0.648 (0.014) & 0.656 (0.014) & 0.674 (0.035) & 0.664 (0.010) \\
Mixed & 0.720 (0.005) & 0.728 (0.005) & 0.627 (0.037) & 0.692 (0.005) \\
\bottomrule
\end{tabular}
\end{table}
\begin{figure}[t]
    \centering
    \includegraphics[width=\linewidth]{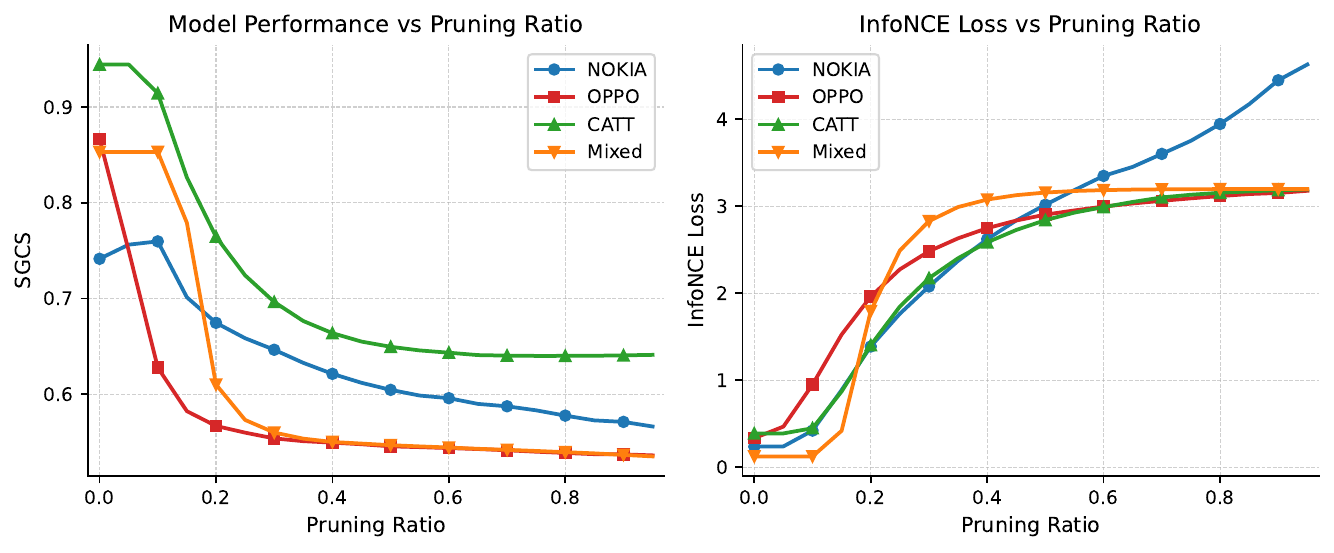}
    \caption{Impact of structured pruning on CPC-before-Compression (GRU, Hidden=128). The left panel reports SGCS ($\uparrow$) and the right panel reports InfoNCE loss ($\downarrow$) as a function of pruning ratio, across all four training datasets. Results are averaged over 5 runs.}
    \label{fig:pruning-analysis}
\end{figure}
    {
    \paragraph{CPC-after-Compression}
    achieves SGCS scores  comparable to the 3GPP baseline while providing future CSI representations at the BS. Scores range from 0.566 to 0.739, with CATT test performance remaining the weakest across all training datasets, consistent with its domain-specific characteristics noted earlier. Notably, InfoNCE loss remains near zero across all configurations (0.005--0.074), suggesting that performing temporal modeling on reconstructed CSI reduces the diversity of latent representations as the information loss happens during compression and decompression. Thus, the contrastive learning loses its effectiveness. As a result, compression bottleneck is the dominant constraint in this variant.
    }    
   { 
    \subsection{Ablation Studies}
{\paragraph{Effect of Temporal Backbone Architecture}
    Fig.~\ref{fig:backbone_ablation} compares GRU, LSTM, and Transformer backbones across both model variants. For CPC-after-Compression, all three backbones converge to nearly identical performance, confirming that the compression bottleneck is the dominant limiting factor rather than the temporal architecture. For CPC-before-Compression, the Transformer achieves marginally higher SGCS, though the gain over GRU and LSTM is negligible for most training datasets. The exception is the NOKIA training dataset, where GRU exhibits higher variance and lower mean SGCS compared to LSTM and Transformer, indicating sensitivity to backbone choice under domain-specific data characteristics. Overall, GRU provides a favorable trade-off between complexity and performance.
    }   }
\begin{figure}[t]
    \centering
    \begin{subfigure}[t]{0.4\textwidth}
        \centering
        \includegraphics[width=\linewidth]{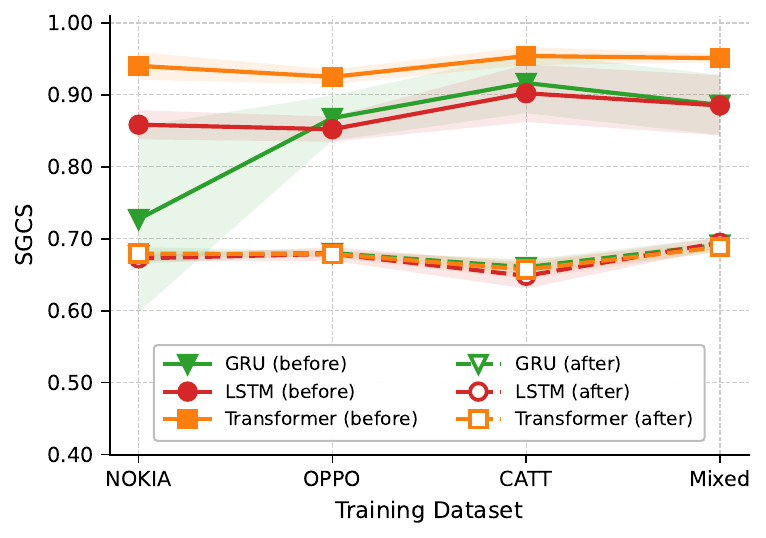}
        \caption{Effect of temporal backbone architecture.}
        \label{fig:backbone_ablation}
    \end{subfigure}\hspace{-0.5em}%
    \begin{subfigure}[t]{0.4\textwidth}
        \centering
        \includegraphics[width=\linewidth]{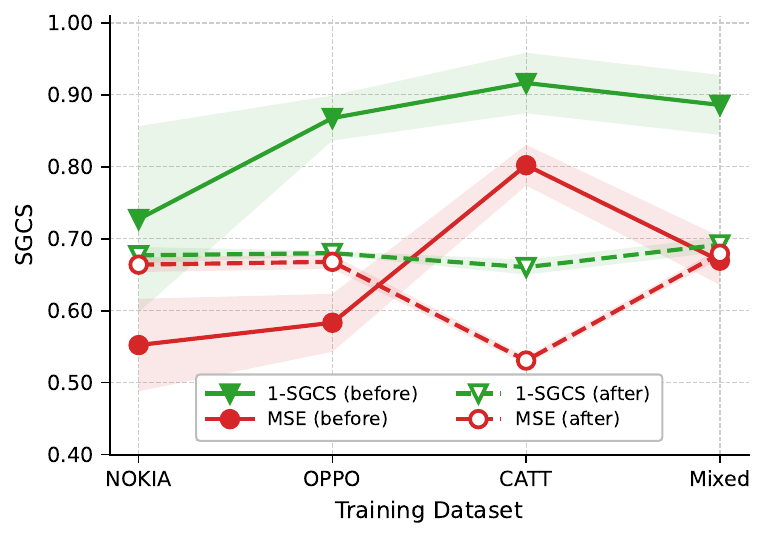}
        \caption{Effect of reconstruction loss (1-SGCS vs.\ MSE).}
        \label{fig:loss_ablation}
    \end{subfigure}
    \caption{Ablation studies measuring SGCS averaged over all test datasets. Solid and dashed lines denote CPC-before-Compression and CPC-after-Compression, respectively. Shaded bands indicate standard deviation over 5 runs.}
    \label{fig:ablations}
\end{figure}
\begin{table}[t]
\caption{Effect of prediction horizon $T$ on SGCS~($\uparrow$) and 
InfoNCE loss~($\downarrow$) averaged over all test datasets and 5 runs.}
\label{tab:pred_horizon}
\centering
\renewcommand{\arraystretch}{1.2}
\resizebox{\columnwidth}{!}{%
\begin{tabular}{ll cccc cccc}
    \toprule
    & & \multicolumn{4}{c}{\textbf{SGCS} ($\uparrow$)} & 
      \multicolumn{4}{c}{\textbf{InfoNCE Loss} ($\downarrow$)} \\
    \cmidrule(lr){3-6} \cmidrule(lr){7-10}
    \textbf{Variant} & \textbf{Train} & 
    $T{=}2$ & $T{=}5$ & $T{=}10$ & $T{=}20$ & 
    $T{=}2$ & $T{=}5$ & $T{=}10$ & $T{=}20$ \\
    \midrule
    \multirow{4}{*}{\shortstack[l]{CPC-before\\-Comp}}
        & NOKIA & $0.704$ & $0.727$ & $0.743$ & $0.778$ 
               & $0.566$ & $0.654$ & $0.938$ & $1.631$ \\
        & OPPO  & $0.834$ & $0.868$ & $0.801$ & $0.891$ 
               & $0.293$ & $0.198$ & $0.421$ & $0.839$ \\
        & CATT  & $0.915$ & $0.916$ & $0.943$ & $0.839$ 
               & $0.514$ & $0.477$ & $0.662$ & $1.568$ \\
        & Mixed & $0.873$ & $0.886$ & $0.817$ & $0.842$ 
               & $0.182$ & $0.203$ & $0.975$ & $1.959$ \\
    \midrule
    \multirow{4}{*}{\shortstack[l]{CPC-after\\-Comp}}
        & NOKIA & $0.677$ & $0.677$ & $0.676$ & $0.675$ 
               & $0.023$ & $0.025$ & $0.019$ & $0.017$ \\
        & OPPO  & $0.680$ & $0.680$ & $0.679$ & $0.675$ 
               & $0.028$ & $0.024$ & $0.022$ & $0.021$ \\
        & CATT  & $0.660$ & $0.661$ & $0.664$ & $0.655$ 
               & $0.017$ & $0.018$ & $0.018$ & $0.018$ \\
        & Mixed & $0.694$ & $0.692$ & $0.698$ & $0.693$ 
               & $0.015$ & $0.013$ & $0.016$ & $0.015$ \\
    \bottomrule
\end{tabular}}
\end{table}
    {   
    \paragraph{Effect of Prediction Horizon}
    Table~\ref{tab:pred_horizon} evaluates the effect of prediction horizon $T$ on 
    SGCS and InfoNCE loss. For \textit{CPC-after-Compression}, SGCS remains nearly 
    constant across all horizons (e.g., NOKIA: 0.677, 0.677, 0.676, 0.675), 
    demonstrating that the model sustains reconstruction quality regardless of how 
    far ahead it predicts. 
    For \textit{CPC-before-Compression}, most training datasets 
    exhibit stable or slightly varying SGCS, with the exception of NOKIA, which 
    improves with longer horizons. This is attributed to NOKIA's stronger temporal 
    autocorrelation structure, where additional prediction steps provide a richer 
    contrastive training signal, enabling the GRU to learn more robust temporal 
    representations. Notably, InfoNCE loss increases with $T$ for 
    \textit{CPC-before-Compression} (e.g., NOKIA: 0.566 at $T=2$ to 1.631 at 
    $T=20$), reflecting a more challenging contrastive objective as the prediction 
    horizon extends, while remaining near-zero for \textit{CPC-after-Compression} 
    across all horizons.}    
    {
    \paragraph{Effect of Reconstruction loss}
    Fig.~\ref{fig:loss_ablation} compares the 1-SGCS and MSE reconstruction losses across both model variants. The 1-SGCS loss consistently outperforms MSE, as it directly optimizes the evaluation metric. The performance gap is particularly pronounced in CPC-before-Compression, where MSE training leads to significant degradation for Mixed, NOKIA, and OPPO training datasets. These results confirm that aligning the training objective with the evaluation metric is critical for effective CSI compression.
    }
\begin{table}[t]
    \centering
    \caption{Effect of compressed representation size on SGCS ($\uparrow$) / InfoNCE ($\downarrow$) averaged over all test datasets and 5 runs. \textbf{Bold} denotes the best SGCS and best InfoNCE among all models for each compression size and training dataset.}
    \resizebox{\columnwidth}{!}{%
    \begin{tabular}{l l ccc}
        \toprule
        \textbf{Model} & \textbf{Train} & \textbf{Size=32} & \textbf{Size=64} & \textbf{Size=128} \\
        \midrule
        \multirow{4}{*}{3GPP Baseline}
            & NOKIA & $0.676$ & $0.704$ & $0.725$ \\
            & OPPO  & $0.673$ & $0.701$ & $0.730$ \\
            & CATT  & $0.653$ & $0.712$ & $0.758$ \\
            & Mixed & $0.691$ & $0.720$ & $0.779$ \\
        \midrule
        \multirow{4}{*}{CPC-before-Comp}
            & NOKIA & $\mathbf{0.727}$ / $\mathbf{0.654}$ & $\mathbf{0.734}$ / $0.756$ & $\mathbf{0.749}$ / $0.617$ \\
            & OPPO  & $\mathbf{0.868}$ / $\mathbf{0.198}$ & $\mathbf{0.833}$ / $0.319$ & $\mathbf{0.794}$ / $0.355$ \\
            & CATT  & $\mathbf{0.916}$ / $\mathbf{0.477}$ & $\mathbf{0.937}$ / $0.620$ & $\mathbf{0.945}$ / $0.622$ \\
            & Mixed & $\mathbf{0.886}$ / $0.203$          & $\mathbf{0.860}$ / $0.422$ & $0.770$ / $\mathbf{0.154}$ \\
        \midrule
        \multirow{4}{*}{CPC-after-Comp}
            & NOKIA & $0.677$ / $\mathbf{0.009}$ & $0.702$ / $0.017$ & $0.743$ / $0.021$ \\
            & OPPO  & $0.680$ / $\mathbf{0.008}$ & $0.708$ / $0.020$ & $0.736$ / $0.021$ \\
            & CATT  & $0.661$ / $\mathbf{0.014}$ & $0.724$ / $0.021$ & $0.771$ / $0.021$ \\
            & Mixed & $0.692$ / $\mathbf{0.005}$ & $0.740$ / $0.015$ & $\mathbf{0.779}$ / $0.017$ \\
        \bottomrule
    \end{tabular}
    }
    \label{tab:bottleneck_ablation}
\end{table}
    {
    \paragraph{Effect of Bottleneck Size}
    Table~\ref{tab:bottleneck_ablation} reports SGCS and InfoNCE across three compressed representation sizes. The 3GPP baseline and CPC-after-Compression improve consistently with larger sizes in SGCS, while their InfoNCE remains near zero across all sizes, confirming that compression is the dominant bottleneck rather than the temporal objective. CPC-before-Compression outperforms both variants at every size, though it exhibits instability at Size=128 for certain training datasets (e.g., $0.770$ for Mixed), suggesting Size=32 offers a better complexity-performance trade-off for the contrastive pre-compression approach.
    }
\begin{table}[t]
    \centering
    \caption{Effect of GRU hidden dimension on model complexity and SGCS/InfoNCE performance for CPC-before-Compression. Results averaged over all test datasets and 5 runs. \textbf{Bold} denotes the best SGCS ($\uparrow$) and best InfoNCE ($\downarrow$) per training dataset.}
    \resizebox{\columnwidth}{!}{%
    \renewcommand{\arraystretch}{1}
    \begin{tabular}{l ccc}
        \toprule
        & \textbf{Hidden=32} & \textbf{Hidden=64} & \textbf{Hidden=128} \\
        \midrule
        Encoder Params   & 663,072    & 2,648,608    & 10,587,168 \\
        Encoder GFLOPs (per batch)   & 0.65              & 2.62              & 10.47               \\
        Enc. Time (ms/batch)   & 1.978                 & 2.0337                  & 2.0984                   \\
        Dec. Time (ms/batch)   & 0.669                 & 0.6801                  & 0.6799                   \\
        \midrule
        \multicolumn{4}{l}{\textit{SGCS $\uparrow$ / InfoNCE $\downarrow$}} \\
        \midrule
        NOKIA & $0.809$ / $0.818$ & $\mathbf{0.823}$ / $1.423$ & $0.727$ / $\mathbf{0.654}$ \\
        OPPO  & $0.826$ / $0.893$ & $0.799$ / $1.009$          & $\mathbf{0.868}$ / $\mathbf{0.198}$ \\
        CATT  & $0.783$ / $2.240$ & $0.903$ / $0.764$          & $\mathbf{0.916}$ / $\mathbf{0.477}$ \\
        Mixed & $0.785$ / $1.251$ & $0.876$ / $0.327$          & $\mathbf{0.886}$ / $\mathbf{0.203}$ \\
        \bottomrule
    \end{tabular}
    }
    \label{tab:hidden_ablation}
\end{table}
{
    \paragraph{Effect of Temporal Hidden Dimension}
    Table~\ref{tab:hidden_ablation} examines the effect of the GRU hidden dimension on model complexity and performance. Larger hidden sizes improve both SGCS and InfoNCE across most datasets, as wider representations enable richer temporal context modeling. Notably, Hidden=128 achieves the best SGCS for OPPO, CATT, and Mixed, while Hidden=64 provides competitive results at a significantly reduced encoder footprint (2,649M vs. 10,587M parameters). Despite the substantial increase in encoder GFLOPs (0.65, 2.62, and 10.47~GFLOPs per batch for Hidden=32, 64, and 128 respectively), per-batch inference latency remains nearly constant (1.978--2.098~ms), as fixed GPU kernel launch overhead dominates short-running kernels~\cite{cuda_graphs_overhead}, further compounded by on-the-fly GRU weight quantization during each forward pass. Overall, Hidden=64 offers a favorable trade-off, achieving over 75\% parameter reduction relative to Hidden=128 with only marginal SGCS degradation.
    }
\begin{table}[bt]
    \centering
    \caption{Effect of decoder capacity on SGCS ($\uparrow$) / InfoNCE ($\downarrow$) for CPC-after-Compression variants. Results averaged over 5 runs.}
    \resizebox{\columnwidth}{!}{%
    \renewcommand{\arraystretch}{1.25}
    \begin{tabular}{l l cccc}
        \toprule
        \textbf{Model} & \textbf{Train} & \textbf{Nokia} & \textbf{Oppo} & \textbf{CATT} & \textbf{Mixed} \\
        \midrule
        \multicolumn{6}{l}{\textit{Decoder: 1,810,754 params \ \ Enc. 2.4457 ms / Dec. 2.9415 ms per batch \ \ Dec. GFLOPs: 1.21}} \\
        \midrule
        \multirow{4}{*}{CPC-after-Comp}
            & NOKIA & $0.730$ / $0.008$ & $0.735$ / $0.008$ & $0.566$ / $0.074$ & $0.677$ / $0.009$ \\
            & OPPO  & $0.733$ / $0.008$ & $0.739$ / $0.008$ & $0.568$ / $0.068$ & $0.680$ / $0.009$ \\
            & CATT  & $0.648$ / $0.014$ & $0.656$ / $0.014$ & $0.674$ / $0.035$ & $0.664$ / $0.010$ \\
            & Mixed & $0.720$ / $0.005$ & $0.728$ / $0.005$ & $0.627$ / $0.037$ & $0.692$ / $0.005$ \\
        \midrule
        \multicolumn{6}{l}{\textit{Decoder: 4,057,026 params \ \ Enc. 2.4648 ms / Dec. 3.6716 ms per batch \ \ Dec. GFLOPs: 3.45}} \\
        \midrule
        \multirow{4}{*}{CPC-after-Comp-Ver2}
            & NOKIA & $0.734$ / $0.036$ & $0.739$ / $0.036$ & $0.568$ / $0.056$ & $0.680$ / $0.034$ \\
            & OPPO  & $0.734$ / $0.024$ & $0.739$ / $0.024$ & $0.569$ / $0.059$ & $0.680$ / $0.024$ \\
            & CATT  & $0.629$ / $0.012$ & $0.635$ / $0.013$ & $0.670$ / $0.018$ & $0.648$ / $0.010$ \\
            & Mixed & $0.729$ / $0.015$ & $0.737$ / $0.014$ & $0.622$ / $0.044$ & $0.696$ / $0.013$ \\
        \bottomrule
    \end{tabular}
    }
    \label{tab:ver2_ablation}
\end{table}
{
    \paragraph{Effect of Decoder Capacity}
    Table~\ref{tab:ver2_ablation} compares CPC-after-Compression with an enhanced variant (Ver2) that introduces additional feature extraction layers before the CPC module at the BS. Since the BS has no strict resource constraints, a larger decoder is practically feasible. Ver2 increases decoder parameters from 1.81M to 4.06M and Dec. GFLOPs from 1.21 to 3.45~GFLOPs per batch, while decoder inference time increases marginally from 2.94 to 3.67~ms per batch. Despite this additional capacity, SGCS improvements are marginal (e.g., NOKIA-trained: 0.730 to 0.734 on Nokia test), and InfoNCE loss increases slightly across most configurations. This suggests that the performance ceiling of CPC-after-Compression is not limited by decoder capacity, but rather by the information loss introduced during the compression-decompression cycle prior to temporal modeling.
    }
            

\section{Conclusion}
This work builds upon the CSI compression model considered in 3GPP and develops two AI-enhanced architectures that integrate contrastive predictive coding to address the problem of channel aging in 3GPP-compliant systems. Through evaluations on 3GPP-compliant datasets from Nokia, Oppo, and CATT, we showed that applying CPC before compression leads to consistently  higher SGCS scores, often exceeding 0.90, and $32\times$ lower decoder GFLOPs compared to the 3GPP baseline. 
{Ablation studies confirmed that the 1-SGCS loss, GRU backbone at hidden size 64, and compression level of 64 bits  offer the best complexity-performance trade-offs. Notably, both variants maintain stable reconstruction quality across prediction horizons of 2 to 20 steps, enabling the BS to obtain future CSI estimates at no additional reconstruction cost.} Additionally, pruning experiments revealed a favorable trade-off between model size and predictive performance~\cite{radwan2025tinyml}.
Future directions include exploring lightweight designs with compressed intermediate representations, and adaptive fine-tuning across channel variations.

\bibliographystyle{IEEEtran}
\bibliography{references}

\vfill

\end{document}